\documentclass[10pt]{article}
\pdftrailerid{}
\usepackage[utf8]{inputenc}
\usepackage[T1]{fontenc}
\usepackage{amsmath,amssymb,dsfont}
\numberwithin{equation}{section}
\usepackage{microtype}
\usepackage{graphicx,tikz,pgfplots}
\graphicspath{{images/}}
\pgfplotsset{compat=newest}
\usepackage[hyperref,amsmath,thmmarks]{ntheorem}
\usepackage{aliascnt}
\usepackage[a4paper,centering,bindingoffset=0cm,marginpar=2cm,margin=2.5cm]{geometry}
\usepackage[pagestyles]{titlesec}
\usepackage[font=footnotesize,format=plain,labelfont=sc,textfont=sl,width=0.75\textwidth,labelsep=period]{caption}
\usepackage{matlab-prettifier}
\usepackage{subcaption}
\newcommand{\lstinlineMatlab}[1]{\lstinline[style=Matlab-editor]{#1}}
\usepackage{alphabeta}

\usepackage[export]{adjustbox}

\usepackage[backend=biber,maxnames=10,backref=true,hyperref=true,giveninits=true,safeinputenc]{biblatex}
\DefineBibliographyStrings{english}{%
	backrefpage = {cited on page},
	backrefpages = {cited on pages},
}

\DeclareFieldFormat[report]{title}{``#1''}
\DeclareFieldFormat[book]{title}{``#1''}
\AtEveryBibitem{\clearfield{url}}
\AtEveryBibitem{\clearfield{note}}

\titleformat{\section}[block]{\large\sc\filcenter}{\thesection.}{0.5ex}{}[]
\titleformat{\subsection}[runin]{\bf}{\thesubsection.}{0.5ex}{}[.]

\usepackage[pdftex,colorlinks=true,linkcolor=blue,citecolor=green,urlcolor=blue,bookmarks=true,bookmarksnumbered=true]{hyperref}
\hypersetup
{
	bookmarksnumbered,
	breaklinks,
	pdfborderstyle=,
	colorlinks,
	citecolor=[rgb]{0,.65,0},
	linkcolor=[rgb]{0,0,.5},
	urlcolor=[rgb]{0,0,.5},
	pdfauthor={Authors},
	pdfsubject={Subject},
	pdftitle={Title},
	pdfkeywords={Keywords}
}

\newpagestyle{headers}
{
	\headrule
	\sethead[\footnotesize\thepage][\footnotesize\sc Authors][]{}{\footnotesize\sc Title}{\footnotesize\thepage}
	\setfoot{}{}{}
}
\newaliascnt{proposition}{lemma}

\aliascntresetthe{proposition}

\newaliascnt{corollary}{lemma}

\aliascntresetthe{corollary}

\newaliascnt{theorem}{lemma}

\aliascntresetthe{theorem}

\theorembodyfont{\normalfont}
\newaliascnt{definition}{lemma}

\aliascntresetthe{definition}

\newaliascnt{assumption}{lemma}

\aliascntresetthe{assumption}

\theoremstyle{nonumberplain}
\theoremseparator{:}
\theoremheaderfont{\normalfont\itshape}

\theoremsymbol{\ensuremath{\square}}

\newcommand{\R}{\mathds{R}}

\newcommand{\RR}{\mathcal{R}}

\newcommand{\call}{\mathrm{l}}

\let\RE\Re
\let\Re=\undefined
\DeclareMathOperator{\Re}{\RE e}
\let\IM\Im
\let\Im=\undefined
\DeclareMathOperator{\Im}{\IM m}

\let\ii\i
\renewcommand{\i}{\mathrm i}

\title{Deciphering scrolls with tomography: A training experiment}
\author{Sonia Foschiatti$^1$\\{\footnotesize\href{mailto:sonia.foschiatti@univie.ac.at}{sonia.foschiatti@univie.ac.at}}
\and Axel Kittenberger$^1$\\{\footnotesize\href{mailto:axel.kittenberger@univie.ac.at}{axel.kittenberger@univie.ac.at}}
\and Otmar Scherzer$^{1,2,3}$\\{\footnotesize\href{mailto:otmar.scherzer@univie.ac.at}{otmar.scherzer@univie.ac.at}}}
\date{}

\begin{document}

\maketitle
\thispagestyle{empty}
\begin{center}
\hspace*{5em}
\parbox[t]{12em}{\footnotesize
\hspace*{-1ex}$^1$Faculty of Mathematics\\
University of Vienna\\
Oskar-Morgenstern-Platz 1\\
A-1090 Vienna, Austria}
\hfil
\parbox[t]{17em}{\footnotesize
\hspace*{-1ex}$^2$Johann Radon Institute for Computational\\
\hspace*{1em}and Applied Mathematics (RICAM)\\
Altenbergerstraße 69\\
A-4040 Linz, Austria}
\end{center}

\begin{center}
\parbox[t]{19em}{\footnotesize
\hspace*{-1ex}$^3$Christian Doppler Laboratory\\
for Mathematical Modeling and Simulation\\
of Next Generations of Ultrasound Devices (MaMSi)\\
Oskar-Morgenstern-Platz 1\\
A-1090 Vienna, Austria}
\end{center}

\begin{abstract}
	The recovery of severely damaged ancient written documents has proven to be a major challenge for many scientists, mainly due to the impracticality of physical unwrapping them. Non-destructive techniques, such as X-ray computed tomography (CT), combined with computer vision algorithms, have emerged as a means of facilitating the virtual reading of the hidden contents of the damaged documents. This paper proposes an educational laboratory aimed at simulating the entire process of acquisition and virtual recovery of the ancient works. We have developed an experimental setup that uses visible light to replace the detrimental X-rays, and a didactic software pipeline that allows students to virtually reconstruct a transparent rolled sheet with printed text on it, the wrapped scroll.
\end{abstract}

\section{Introduction}

In recent years there has been a growing interest in the recovery of ancient written documents that have been damaged by time, inadequate storage conditions, chemical corrosion, and misguided attempts to open them. For instance, papyrus sheets, produced from the papyrus plant, were a widely used writing surface for documents in antiquity. However, the surviving scrolls and fragments require specific climatic conditions for their preservation. High levels of humidity can cause severe deformations in their layers, altering the written content. In such cases, physical unwrapping is not feasible as it increases the risk of permanent damage. Moreover, for other materials such as parchment, wood, and bamboo, the physical cleaning processes can lead to material degradation and require a substantial time investment. 

Recent advancements in the field of computer science, particularly in the areas of image processing and computer vision, have paved the way for virtual restoration of damaged documents avoiding physical unwrapping, a process called \textit{virtual unwrapping}. The term "virtual unwrapping", adopted from \cite{SeaParSegTovSho16}, refers to the use of software to reveal on a computer screen words and texts hidden for centuries, thereby making the content accessible to all. Virtual reconstructions are based on data acquired through various tomography modalities. The term "tomography" derives from the Greek word "tomos", meaning "slice", and can be defined as a method of imaging by sections using any type of penetrating wave. The employment of X-ray tomography in the recovery of ancient manuscripts can be attributed to the non-destructive and non-invasive nature of X-ray radiation. Its application to paleontology and archaeology can be traced back to the early 20th century, as evidenced by the use of these imaging techniques to reveal ancient texts buried in altars, or rolled silver scrolls, sealed letters, damaged manuscripts from ancient libraries, and carbonized scrolls that have been discovered in the Villa of the Papyri near Herculaneum, Italy \cite{MocBruFerDel15}.

In this note, we present a didactic experiment for the virtual restoration of text printed on a transparent scroll. Drawing inspiration from the intriguing archaeology application, we have developed an educational laboratory for teaching X-ray tomographic reconstruction of written documents. This initiative is particularly well-suited for undergraduate courses in applied mathematics, physics, or archaeology, as well as Maker Fairs or science communication workshops. As previously stated, X-ray tomography is a well-established imaging technique that, in the field of archaeology, allows scientists to read hidden text without damaging the actual document. In our experimental setup, we employ visible light as a surrogate to simulate the potentially detrimental X-rays. This approach is based on the premise that the propagation characteristics of electromagnetic waves, despite their different wavelengths, can be described by analogous mathematical formalism. Within our experimental framework, the wrapped scrolls are composed of transparent cellulose acetate sheets, which exhibit minimal interaction with visible light. The idea of employing a transparent film and the light as an illumination source to simulate X-ray tomography in an educational setting was initially proposed by Kittenberger et al. \cite{MR4416990}.

\begin{figure}[h]
	\centering
	\includegraphics[width=.5\linewidth]{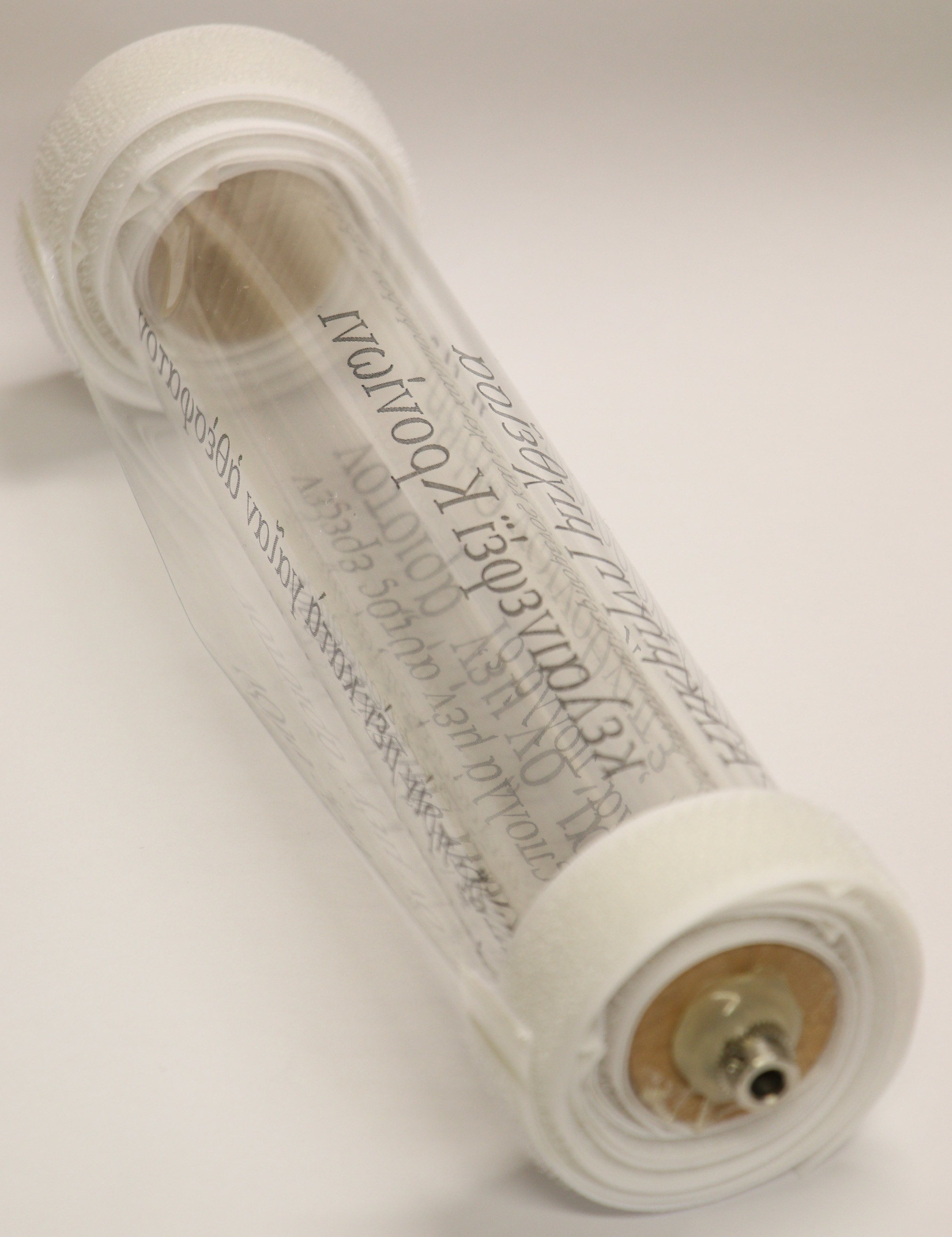}
	\caption{The Greek phantom scroll.}
	\label{fig:greek}
\end{figure} 

The present educational experiment further extends the ideas outlined in \cite{MR4416990}, using a setup comprising a light source, a stepper motor and a digital camera. To simulate the ancient writings, a Greek text from Homer's Hymni Homerici XV (VII-VI century BC) and a Latin text from Lucrezio's De Rerum Natura have been printed on a transparent film (see Figure \ref{fig:greek}). The objective of this experiment is twofold: firstly, to stimulate undergraduate students' or amateurs' interest  in the fascinating field of applied mathematics, which intersects pure mathematics, computer science, numerical analysis and physics; and secondly, to provide instructors with an inspiring application of X-ray tomography that can be used to develop students' problem-solving abilities and scientific thinking.

The Archeolab website (\url{https://csc1.gitlab.io/archeolab/}) has been developed as a virtual guide to assist students and instructors in the reconstruction of their own scrolls. It provides documentation of the Python and Matlab code, in addition to the theoretical foundations.

\section{X-ray tomography in a nutshell}\label{sec: model}

X-ray computed tomography (abbreviated as X-ray CT) is a non-destructive imaging technique used to visualize the internal structure of an object. The key observation granting the success of this approach is that different materials exhibit differed absorption and scattering properties when interacting with X-rays. They pass through the majority of materials (e.g., most of the objects are transparent to them), enabling the clear identification of internal structures of interest through contrast imaging and algorithms. From a physical point of view, in the electromagnetic spectrum X-rays are characterised by shorter wavelengths than visible light, ranging between 10 nanometers and 10 picometers. 

From a mathematical perspective, the problem of recovering the attenuation coefficients of an object is formulated as the problem of determining a function from the knowledge of its line integrals. The first solution to this problem was proved by Johann Radon, an Austrian mathematician who derived the first reconstruction formula in his 1917 paper \cite{MR1313944}.

The first documented application was by Bracewell in radioastronomy in 1956 \cite{MR80017}, who sought to reconstruct a map of the solar microwave emissions from a series of radiation measurements across the solar surface. In the same years, the the South African-American physicist Allan M. Cormack made a significant contribution to the field of cancer treatment by radiation therapy. In 1955, he spent a day and a half a week at Groote Schuur Hospital in Cape Town, South Africa. During this period, he observed the planning of radiotherapy treatments, and recognized the importance of determining the attenuation coefficient of the object. During his sabbatical year at Harvard in 1956, he derived a mathematical theory for image reconstruction. Subsequent to his return to South Africa in 1957, he proceeded to test his theory on a 5-cm-thick disk measuring 20 cm in diameter composed by different materials in order to simulate the skull. Thanks to its radial symmetry, only one measurement at one angular position was needed, and the attenuation coefficient was calculated applying the formulas derived by Allan M. Cormack.

The development of the first clinical CT scanner was initiated in 1967 by Godfrey N. Hounsfield at the Central Research Laboratories EMI in England. Hounsfield, independently of Cormack, deduced that X-ray measurements of a body taken from different directions would allow the reconstruction of its internal structure. In 1979 Allan M. Cormack and Godfrey N. Hounsfield were jointly awarded the Nobel Prize in Physiology and Medicine for their seminal contributions (see \cite{Hsi22} for the complete history).

In the field of archaeology, the detrimental X-rays are employed to examine scrolls without the need for invasive physical interaction, thereby preserving their integrity. An X-ray experiment comprises a source, a detector, and the object to be imaged. During the scanning process, a source emits one or more thin X-ray beams, modelled as straight lines. When the beams encounter the object, their intensity decreases as they interact with its electrons. Detectors placed opposite the sources measure the intensity diminution. In CT, each X-ray image captured at a fixed angle is called a projection. A series of projection data is acquired by rotating the X-ray sources and detectors around the body, which is motionless. By acquiring a sufficient number of projections, it is possible to reconstruct multiple cross-sections, or slices, of the three-dimensional object, resulting in a volumetric representation of the body called volumetric data or volumetric image. The information regarding location and intensity is saved in a single unit called a \textit{voxel}, which represents the three-dimensional counterpart of the two-dimensional pixel. The slices represent a map of the spatial distribution of the attenuation coefficient of the object.

The process of extracting hidden text from volumetric X-ray CT scans of the objects of interest using computer vision algorithms is often referred to as a \textit{software pipeline} or \textit{workflow}. To illustrate this process, we will refer to a successful application of the virtual pipeline, the digital reconstruction of the En-Gedi scroll presented in \cite{SeaParSegTovSho16}. In 1970, several fragments of carbonized parchment were unearthed at the archaeological site of En-Gedi, west of the Dead Sea, in an altar. These fragments are pieces of charcoal, some of which have undergone disintegration and deterioration due to human interaction and other agents over time. The data collected are volumetric X-ray micro-CT scans which provide high-resolution greyscale cross-sections of the body. These images facilitate the distinction of the internal layers of the scroll. The team led by Professor W. B. Seales developed a virtual pipeline with the objective of recovery the concealed text of the scrolls without altering them. The full process consists of the following four steps: segmentation, flattening, texturing, and merging. The initial step, segmentation, involves the identification of layers in each slice of the volume (e.g., the isolation of thin, warped regions of interest). This operation poses a significant challenge due to the presence of noisy measurements in the scans derived from the animal's skin. The result of this initial phase is a segmented surface representing a layer of the scroll. The subsequent step is texturing, which consists in assigning the intensity values to the corresponding points on the segmented surface. The subsequent step is flattening, and consists in mapping the three dimensional wrapped segmented surface onto a two-dimensional image. Finally, the different segments of the flat surface obtained from the slices are merged to create an image of the entire readable scroll. 

Virtual extraction of hidden information often requires user interaction during the segmentation step, a process that could be time-consuming in the long term. A novel approach to overcome this difficulty involves automated segmentation in conjunction with Machine Learning algorithms. In March 2023, the same research group initiated a challenge to develop a fully automated virtual pipeline to disclose ancient carbonized scrolls \cite{VC}. The \textit{Vesuvius Challenge} is a Machine Learning and computer vision competition, with the objective of developing an automated software pipeline capable of reading the hidden text of ancient carbonized scrolls discovered near Herculaneum, Italy. Let us travel back in time to the thriving years of Roman empire when a cataclysmic event happens. In 79 AD there was a devastating eruption of Mount Vesuvius; the towns of Herculaneum and Pompeii along with their buildings and palaces were covered by pyroclastic flow, reaching depths of up to 30 meters. Among them was the villa of Lucio Calpurnius Piso Pontifex, the father-in-law of Julius Caesar, buried under the ashes. The luxurious villa, known as the Villa of the Papyri, contains the only complete library of the Greco-Roman world that has survived to the present day. For centuries, the villa and the other building remained noiselessly concealed beneath the ashes, their existence obscured, until the mid-eighteenth century. Around the year 1750 AD, an Italian farmer, while digging a well, uncovered marble pavements, opening the way to possibly groundbreaking archaeological discoveries. The first excavations on the site were conducted by the Swiss architect Karl Weber between 1750 and 1765, employing tunnels. During these excavations around 1,800 carbonized scrolls were disinterred, referred to as the \textit{Herculaneum papyri}. Ingenuous attempts to open them caused the irremediable loss of ancient books along with their content. It seemed that the text of those manuscripts would have been undisclosed for centuries. Fortunately, the advancements in science and technology have allowed us to aspire to reveal once and for all the concealed text of the Herculaneum papyri. The same group led by Professor Seales conducted X-ray data acquisition from 2010 to 2020 using different X-ray CT machines, collecting the volumetric images in the EduceLab dataset. The acquired data encompasses two types of imaging: besides the volumetric X-ray micro-computed tomography, employed to image the entire collection (e.g., intact scrolls and detached scroll fragments), it includes spectral imaging of surface fragments, used to detect the presence of exposed ink on the surface \cite{par2024}. 

Many amateurs and scientists have participated in the challenge, which has captured the attention of the media and the general public. In 2024, a significant discovery was made by the 21-year-old computer scientist Luke Farritor, who was able to identify the word $\pi\omicron\rho\varphi\nu\rho\alpha\zeta$ meaning "violet" (color or tissue). Another recent discovery from the Herculanei scroll is a newly deciphered passage from a scroll written by Philodemus of Gadara, which reveals details about the lost tomb of the Greek philosopher Plato \cite{DS}. It is still possible to take part in the challenge, and rich prizes are awarded each month for relevant contributions to the software pipeline.

\section{Experimental setup}\label{sec: exp}

The groundbreaking discoveries and achievements that have followed the development of X-ray CT have motivated us to develop a didactic laboratory that allows students and amateurs to tackle the same problem that Johann Radon, Allan M. Cormack and Godfrey N. Hounsfield solved with a focus on the archaeological application.

In the preliminary part of the laboratory, the acquisition of X-ray projections of an ancient written document is simulated. In out setup, we consider two phantoms, representing the scrolls, composed of thin cellulose acetate (CA) sheets, each of size 21 cm x 15 cm. To simulate the text of Herculanean scrolls, Greek and Latin texts have been selected for printing. The first phantom features text from \textit{Hymni Homerici XV - In Herculem}, verses 1-3:

\begin{center}
	Ἡρακλέα Διὸς υἱὸν ἀείσομαι, ὃν μέγ' ἄριστον γείνατ' ἐπιχθονίων Θήβῃς ἔνι καλλιχόροισιν
	Ἀλκμήνη μιχθεῖσα κελαινεφέϊ Κρονίωνι
	
	ὃς πρὶν μὲν κατὰ γαῖαν ἀθέσφατον ἠδὲ θάλασσαν πλαζόμενος πομπῆισιν ὕπ᾿ Εὐρυσθῆος ἄνακτοςπολλὰ μὲν αὐτὸς ἔρεξεν ἀτάσθαλα, πολλὰ δ᾿
	
	ἀνέτλη·νῦν δ᾿ ἤδη κατὰ καλὸν ἕδος νιφόεντος Ὀλύμπουναίει τερπόμενος καὶ ἔχει καλλίσφυρον Ἥβην.
\end{center}

The second phantom features text from \textit{De Rerum Natura} by Lucretius, verses 72-74:

\begin{center}
	
	Ergo vivida vis animi pervicit, et extra
	
	processit longe flammantia moenia mundi
	
	atque omne immensum peragravit mente animoque
	
	unde refert nobis victor quid possit oriri,quid nequeat, finita potestas denique
\end{center}

The original texts can be found in \cite{homer_hercules}.

\begin{figure}[h]
	\centering
	\includegraphics[width=.5\linewidth]{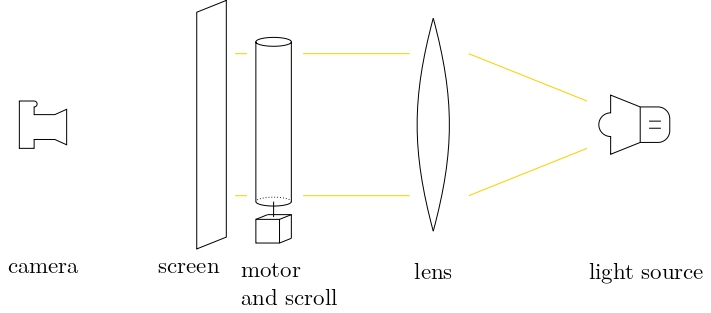}
	\includegraphics[width=.5\linewidth]{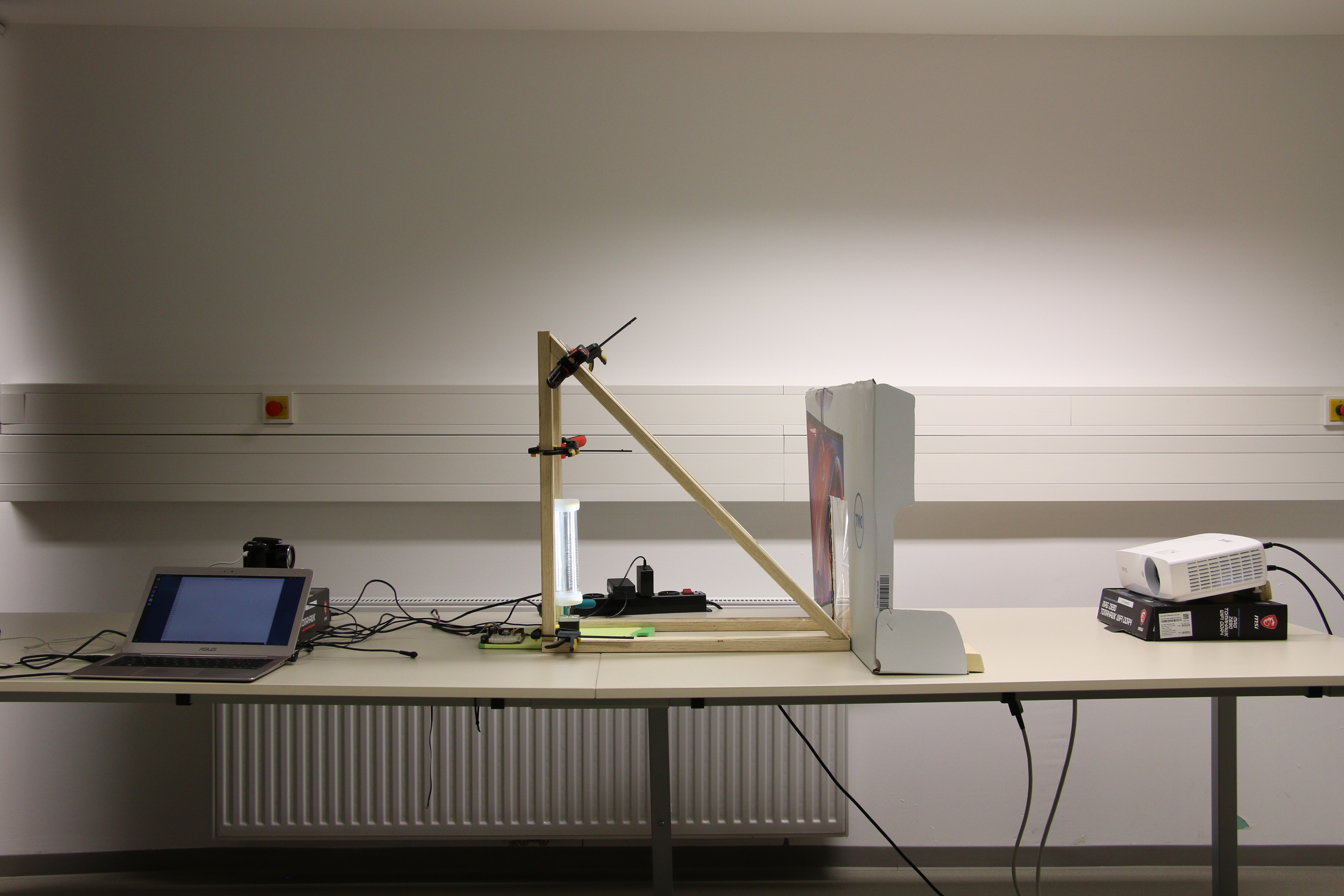}
	\caption{The experimental setup (top: schematic, bottom: our build).}
	\label{fig:experimentalSetup}
\end{figure} 

The experimental setup consists of an Arduino, a breadboard, a computer, a digital camera, a Fresnel lens, hook-and-loop tape, a paper sheet, a projector, a stepper motor, and mounting structures for the lens and paper sheet. In this case, the structures were constructed using cardboard and wooden beams (see Figure \ref{fig:experimentalSetup}).
To avoid fused layers in the reconstruction, the hook-and-loop tape is applied to the phantoms, which are then wrapped. The phantoms are mounted on a pulley, which is screwed onto the axle of a NEMA 17 stepper motor. This is either glued to a base or clamped in a vice to provide a stable vertical axis of rotation, oriented orthogonally to the stage. The complete setup is illustrated in Figure \ref{fig:experimentalSetup}.

The structure is illuminated by a projector, with a 30 cm x 30 cm PMMA Fresnel lens placed at a distance determined by the lens's focal length (60 cm in this case). The use of a Fresnel lens serves to mitigate the parallax effect, which otherwise leads to the formation of significant artifacts in the reconstruction. A white paper acts as the canvas for the experiment. On the opposite side of the canvas, a digital camera is positioned as the detector. A USB cable connects a personal computer and the board controller, with the Arduino controlling the motor stepper via an A4988 driver. The computer controls the Arduino and the digital camera, which is connected to it via a USB cable, using the Canon Hack Development Kit (CHDK) \cite{CHDK}. All the relevant scripts can be found at \url{https://gitlab.com/csc1/archeolab/}.  

The experiment starts with the acquisition of the projection data. The CA phantoms are rotated at a frame rate of 0.44 fps around 360$^{\circ}$, for a total of 30 minutes for 800 shots. The total number of projections acquired is $N=800$. In this particular instance, the remote-shoot function of the CHDK is used to transfer the images to the PC.

\begin{figure}[h]
	\centering
	\includegraphics[width=.5\linewidth]{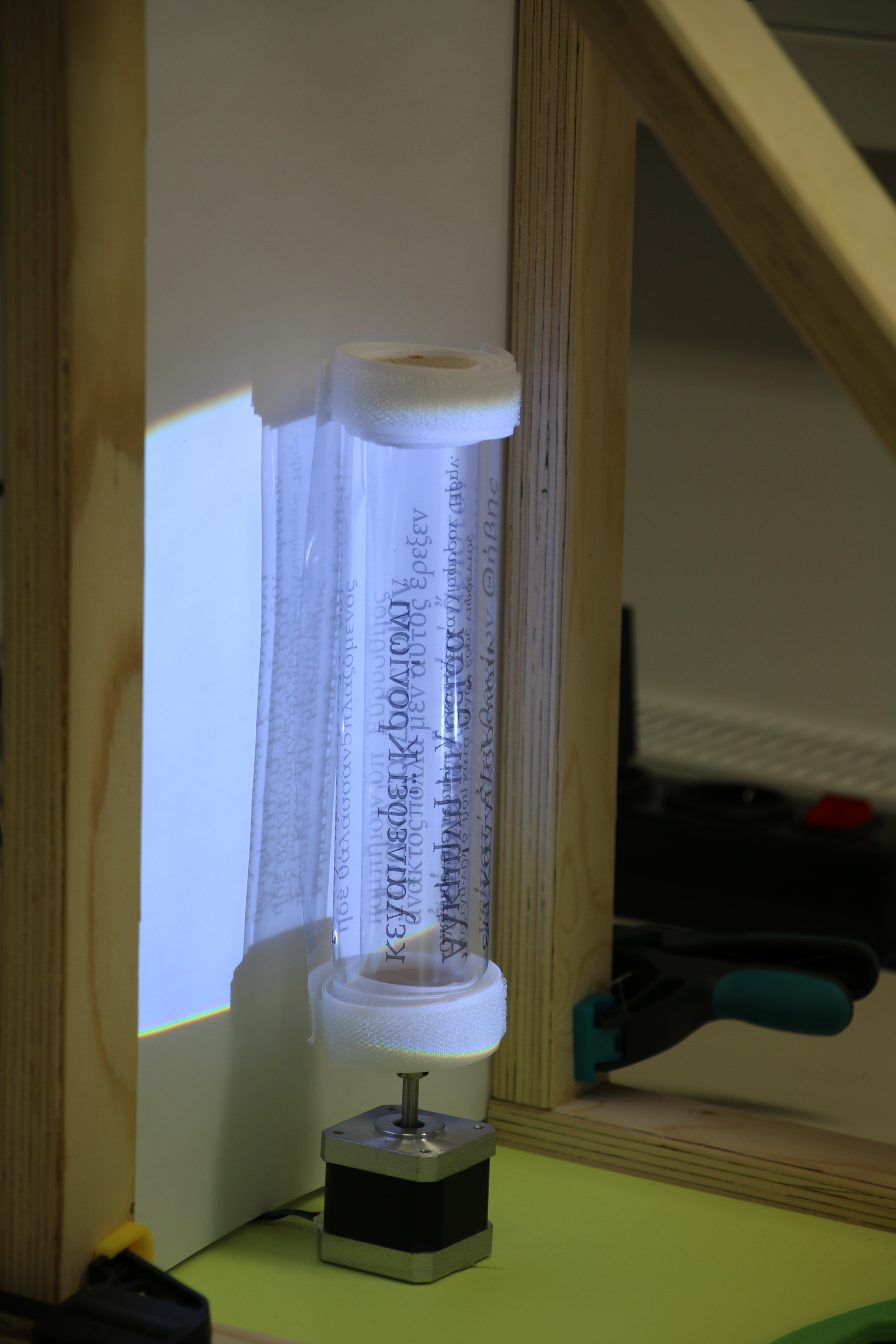}
	\caption{Projection on the white screen.}
	\label{fig:greek_scroll}
\end{figure}

\section{The virtual unwrapping}\label{sec: virtual}

In the subsequent stage of the practical laboratory session, students will be instructed in the utilization of the virtual pipeline employed for the reconstruction of the contents of the transparent scrolls. The main algorithms employed in this process are the Filtered Back Projection algorithm and the Maximum Intensity Projection (MIP) method. The only manual step required by the user is the definition of a set of control points, termed 'individuals', describing a first rough approximation of the spiral-shaped layer of the scroll, which is subsequently optimized to obtain the best fit. 
The measurements are derived from the projections of the transparent scroll on the canvas (see Figure \ref{fig:greek_scroll}).

The virtual workflow commences with the determination of the axis of rotation of the scroll with the call of the script \lstinlineMatlab{guiAxisFinder}. After loading the projections, the GUI (graphical user interface) takes the first image at index $i$ and its 180° counterpart. Subsequently, the images undergo a series of pre-processing steps (i.e., cropped based on a central point, rotated and cut). In the ideal case, where the center and the axis of rotation are configured correctly, the first image and the second flipped should overlap perfectly, resulting in a uniformly black image. However, in practice, the user observes green and red regions in the plotted image, which correspond to negative and positive pixel intensity differences between the two images, respectively. The process of minimizing these differences serves to converge to the actual rotation axis.

\begin{figure}[h]
	\centering
	\includegraphics[frame,width=0.5\linewidth,trim={65mm 10mm 55mm 80mm},clip]{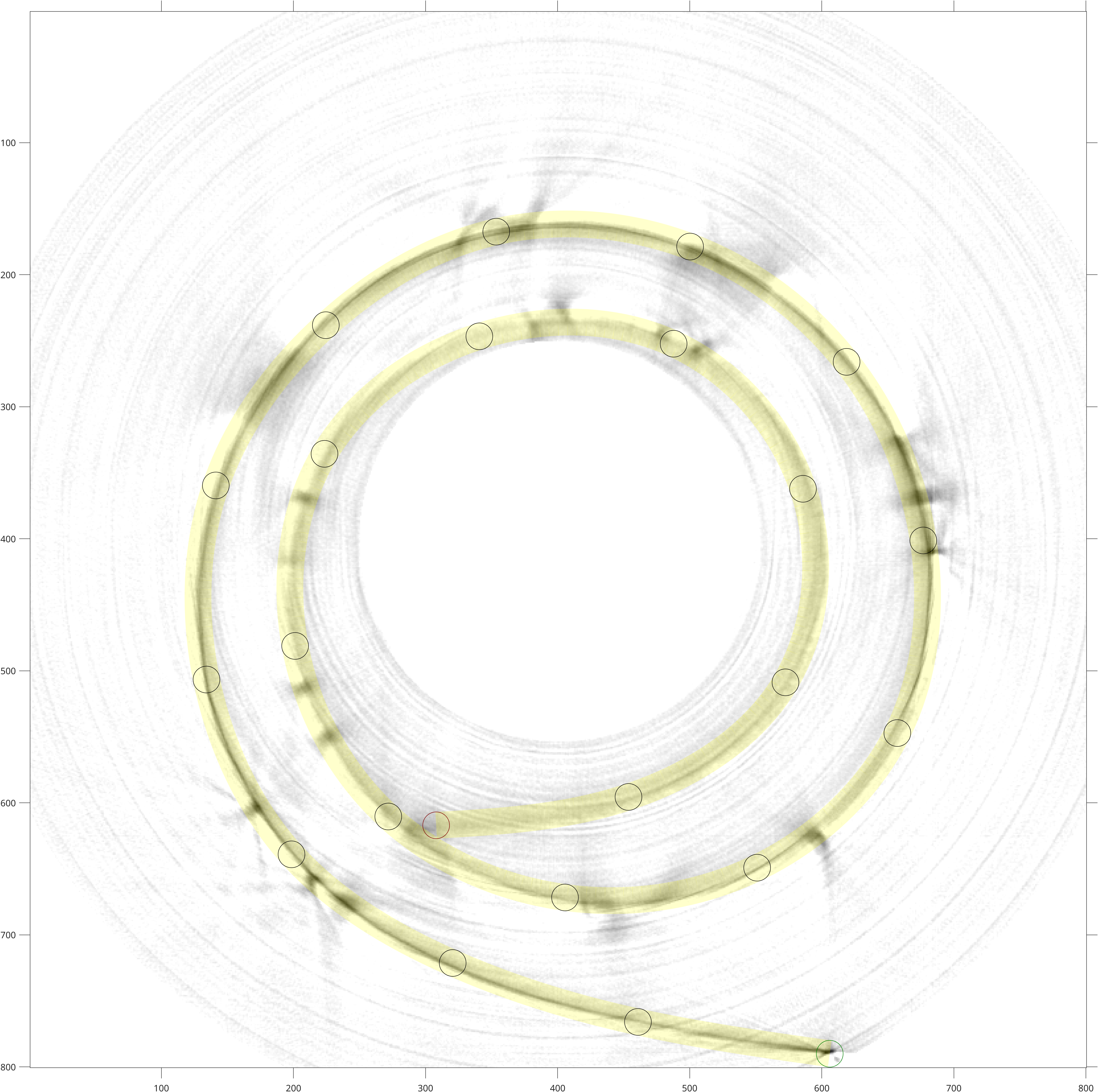}
	\caption{Cross-section of the transparent scroll (inverted) with a segmented path.}
	\label{fig:slice}
\end{figure}

Following this step, the projections are associated with the correct rotation axis, loaded and elaborated to obtain a stack of pre-processed projections. At this point, Filtered Back-Projection is performed, with the stack of pre-processed projections from the previous step serving as the input. The \lstinlineMatlab{iradon} function is then applied with linear interpolation and the Ram-Lak filter as options, which are more suitable for edge detection. The \lstinlineMatlab{iradon} function in MATLAB is available in the Image Processing Toolbox, in Python in the \lstinlineMatlab{skimage} library. The result of this step is a stack of cross-sections of the original scroll, where the layers of the scroll can be seen (see Figure \ref{fig:slice}). After Filtered Back-Projection, a mask is implemented to remove artifacts from the central and external regions of the slices.

Subsequently, the script \lstinlineMatlab{guiSegment} is invoked to perform manual segmentation. The user then inserts a set of initial control points on a selected cross-section of the scroll, drawing a spiral-shaped curve that follows the layers of the scroll. A genetic algorithm is then employed to optimize the manual segmentation. Subsequently, a cubic smoothing spline through the control points is calculated and sampled to obtain a set of points for plotting.

\begin{figure}[h]
	\centering
	\includegraphics[width=.2\linewidth]{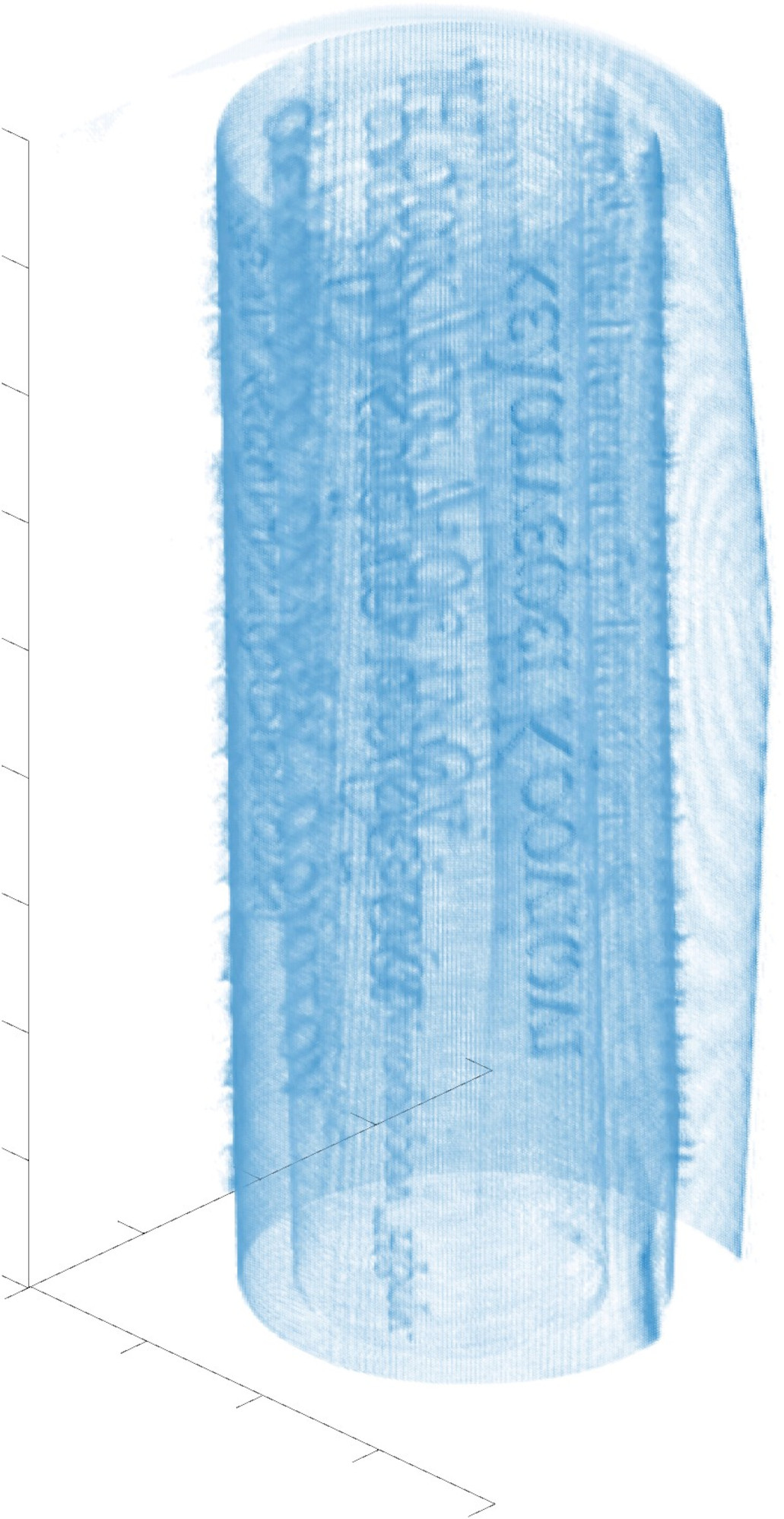}
	\caption{3-D view of the Greek phantom scroll.}
	\label{fig:view}
\end{figure}

\begin{figure}[h]
	\centering
	\includegraphics[frame,width=.3\linewidth]{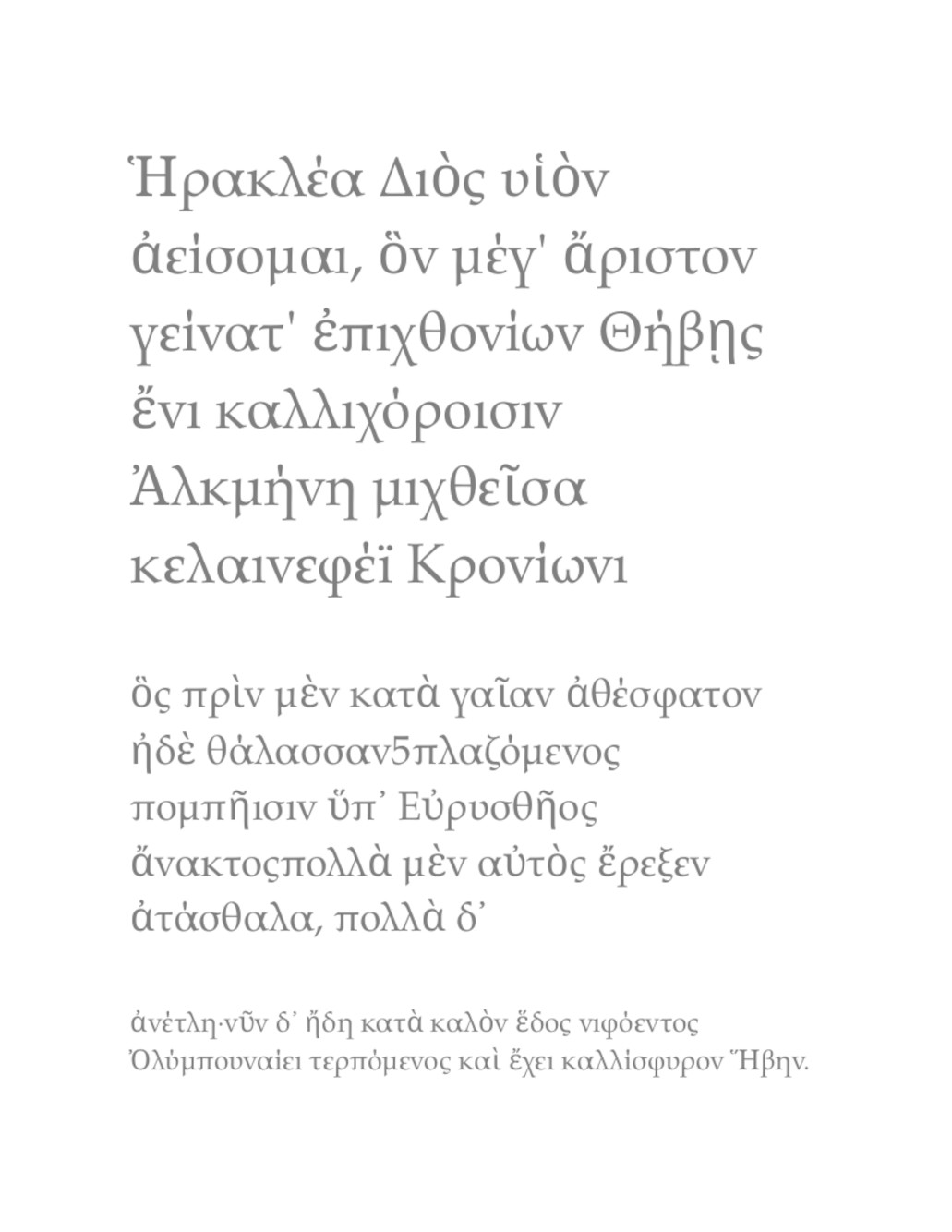}%
	\includegraphics[frame,width=.3\linewidth, clip,trim=0 72 0 52.5]{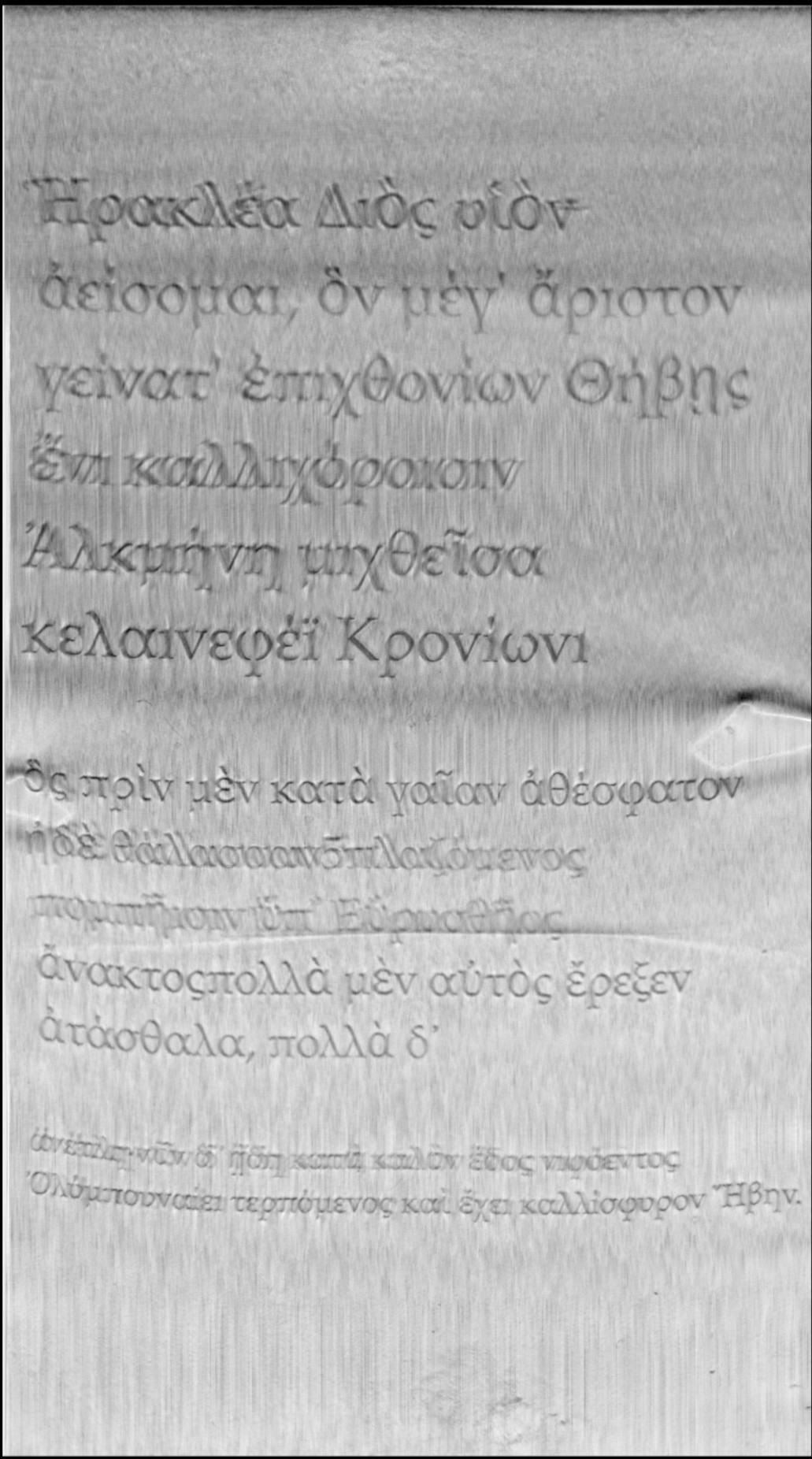}
	\caption{Reconstruction (right, inverted) of the Greek scroll (left).}
	\label{fig:greek_recon}
\end{figure}

\begin{figure}[h]
	\centering
	\includegraphics[frame,width=.3\linewidth]{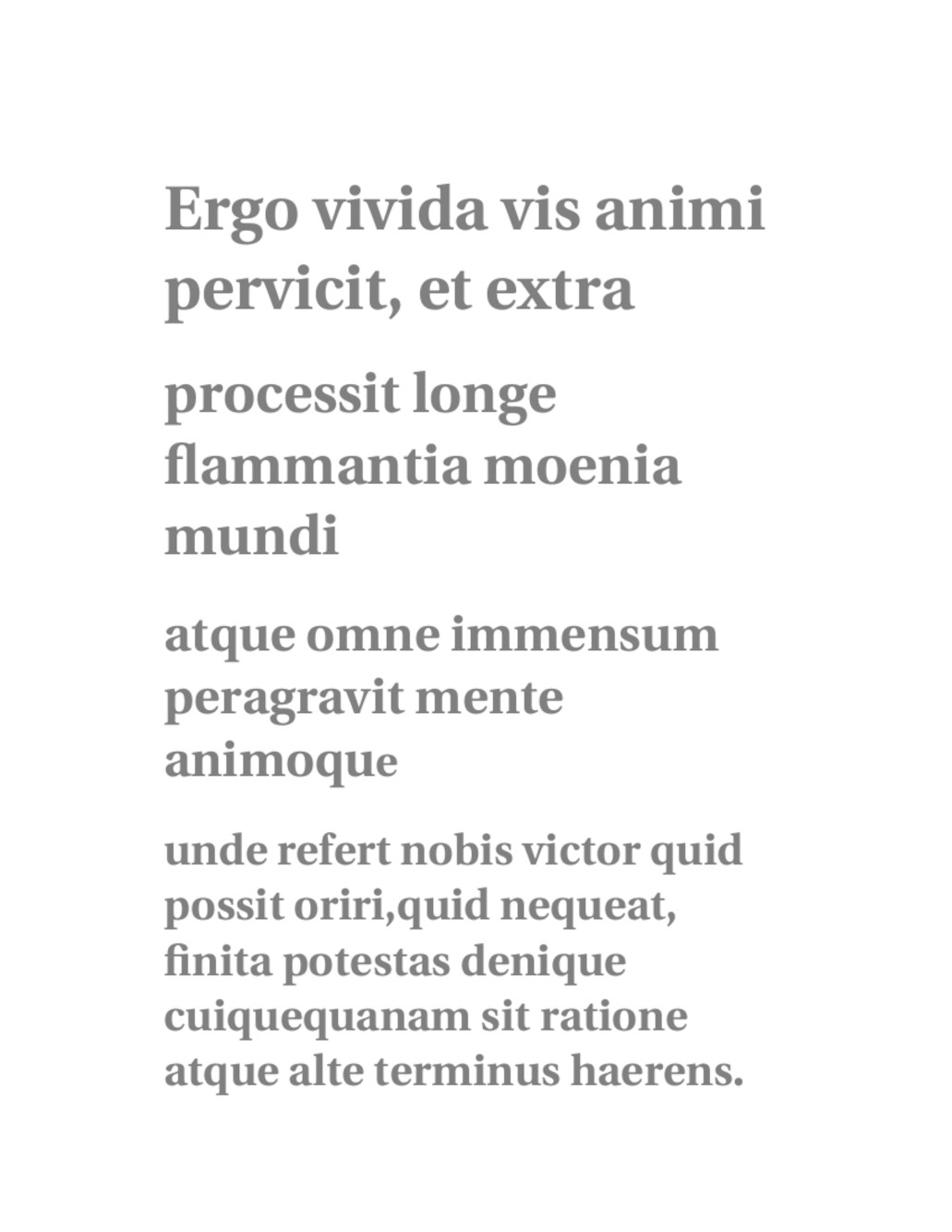}%
	\includegraphics[frame,width=.3\linewidth,clip,trim=0 59 0 50]{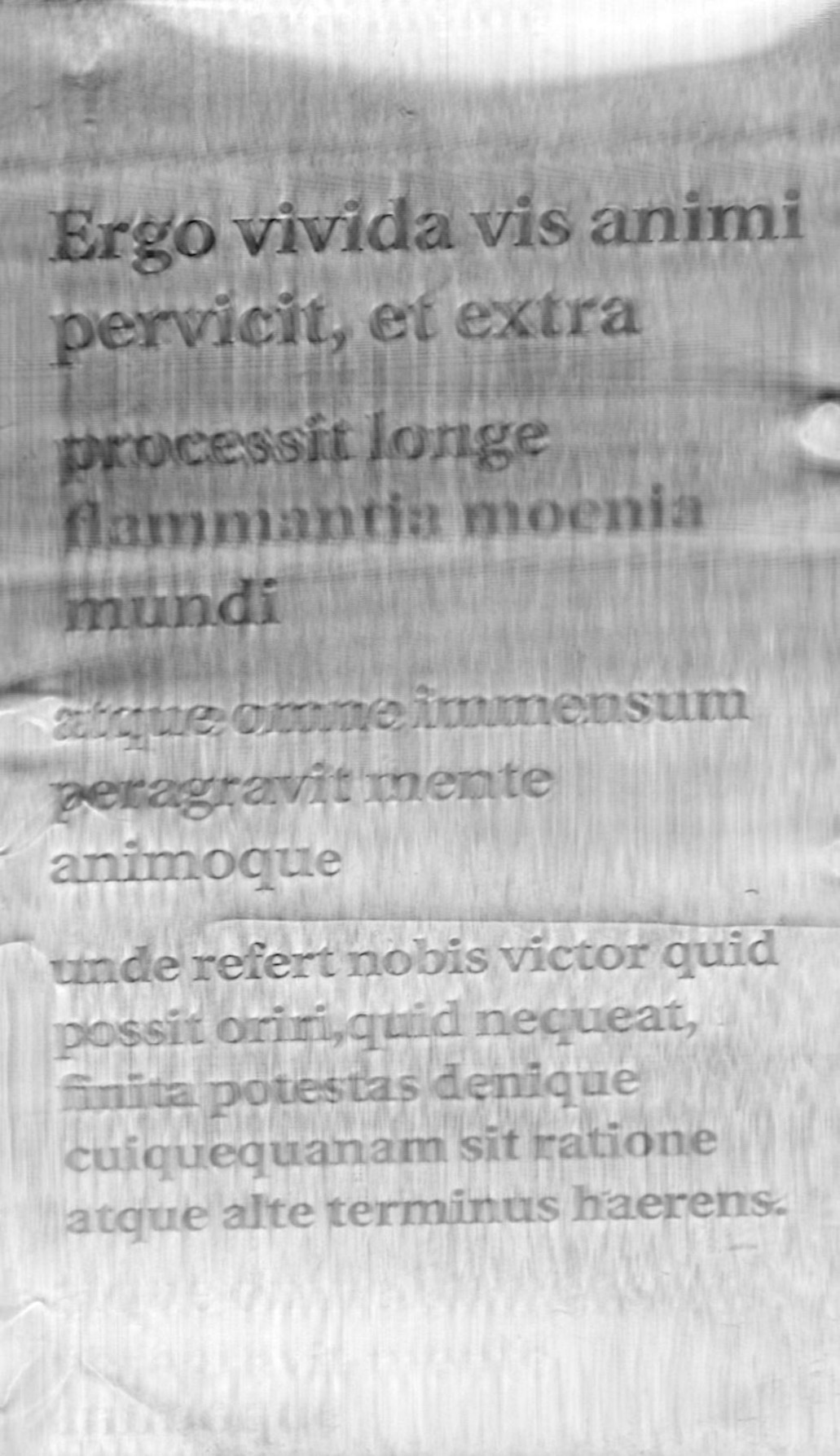}
	\caption{Reconstruction (right, inverted) of the Latin scroll (left).}
	\label{fig:latin_recon}
\end{figure}

The final step involves the implementation of the Maximum Intensity Projection (MIP) method, which reveals the hidden text of the scroll. The MIP method selects the maximum intensity value $I(x,y,z)$ from the voxels $V(x,y,z)$ for a fixed pair of coordinates $(x,y)$:
\begin{equation}
	MIP (x,y) = \max_z I(x,y,z).
\end{equation}
This method projects the maximum intensity values onto a plane, thereby facilitating the visualization of the text. The reconstruction of the Greek and Latin transparent scroll is shown in Figure \ref{fig:greek_recon} and \ref{fig:latin_recon}, and a three dimensional view of the Greek scroll in Figure \ref{fig:view}. Despite the presence of artifacts, most of the Greek and Latin letters are legible without the application of any post-processing steps. 

As illustrated in the figures, characters of greater dimensions exhibit less blurring compared to smaller ones. This disparity can be attributed to the location of words in the internal layer, which may have been more susceptible to interference from the external writings during the reconstruction process.

\section{Laboratory Description}

The material presented in this note may be adapted for a demonstration intended to undergraduate students enrolled in courses related to applied mathematics or physics. The educational experiment has been designed to develop and strengthen the mathematical intuition of tomography modeling, a process that necessitates active student participation. The experimental phase of the activity can be structured in two formats: a group activity or as an instructor demonstration. 

In the first scenario, students are tasked with constructing part of the setup using provided instructions and recording the projections. It is recommended that the instructor undertake a preliminary trial of the experiment. It is strongly advised that the experiment is performed in advance, as this will provide a general overview of the materials and time required to prepare it. Furthermore, it is essential that the instructor is familiar with the code, which can be incorporated into the course's theoretical material. Within the group framework, each team can be assigned a transparent phantom containing either the same sentences or different ones to introduce more variety in the results. 

In the second framework, the instructor will conduct the experiment, with the assistance of a laboratory technician if needed to facilitate the setup assembly. The acquisition of the projections is expected to take approximately 30 minutes, during which time the instructor may introduce the mathematical model of X-ray tomography or explain the virtual pipeline and the code. A concise overview of the mathematical model of X-ray with a summary of the relevant notions is provided in the last section of this contribution. The documentation of the virtual pipeline is available in the form of a Matlab workflow or a Jupyter Notebook on the webpage \url{https://gitlab.com/csc1/archeolab/}. Subsequent to the acquisition of the projections, the students will utilize the code and adhere to the implemented workflow, adjusting the parameters and control points as required.

Following the completion of the virtual unwrapping, students may be requested to write a report or a presentation that summarizes the results obtained, the observations made, the difficulties encountered, and possible improvements. The report could be structured like a scientific article including the following sections: Introduction, Methods, Results, Discussion and Conclusion. The Introduction should provide an overview of the experiment and its objectives. In the Methods section, students should detail the experimental setup and the algorithms that were utilized. The subsequent section is dedicated to the description of the results, including the reconstructed scrolls. The final sections summarizes the experimental outcomes, the problems experienced, and gives suggestions for potential improvements and future directions.

\section{Expected Results and Conclusion}\label{sec: conclusion}

The proposed training experience serves as a valuable resource for instructors in addressing the complexity of the subject matter as well as in the abstract modeling of physical phenomena. 

We proceed to analyze the potential outcomes of the two frameworks described. From a didactic perspective, we anticipate significant interaction among students and instructors, with the latter posing questions and exploring diverse methodologies. The presentation component or the final report constitutes a pivotal element of the laboratory, offering valuable feedback to the instructor concerning the students' comprehension of the subject and the evaluation of their problem-solving skills.

The reconstruction of the text is deemed successful when the resulting images clearly display the text, with the majority of letters being legible (see Figures \ref{fig:greek_recon} and \ref{fig:latin_recon}). Characters within the internal layers are expected to be more blurred compared to the external layers, due to light bending effects. The cross-sections obtained through the reconstruction process are expected to reveal the layers of the phantom, thereby enabling students to comprehend the structure and depth of the text. It is expected that not all groups will possess the capability to fully reconstruct the transparent scroll. Therefore, a database containing tuned images is included to serve as a model for the expected reconstruction and to assist groups lacking this capability in performing the virtual reconstruction.

In the second framework, upon the completion of the experiment by the instructor, students are tasked with processing the projections and constructing their own reconstructed text. Also in this case, the report or presentation will reflect the assimilation of the theory, and outline the encountered challenges, the results obtained, and the mathematical model employed. Indeed, at the end of the experimental laboratory, students should be able to provide clear explanations of the experiment, results obtained, and ability to connect these to the theoretical framework. From a theoretical standpoint, it is expected that students will grasp the fundamentals of X-ray tomography and its significance in the domain of archaeology. In both workshops, students have the opportunity to sharpen their problem-solving skills, whether individually or collectively. Moreover, the final discussion will contribute to promote peer instruction, an effective learning methodology substantiated by current educational studies.

In summary, this note has proposed a simulated recovery of hidden text in ancient papyrus scrolls, utilizing an educational setting to illustrate the principles of X-ray tomography and the development of a rudimentary virtual pipeline for hidden text recovery. It provides detailed instructions for constructing the experimental setup and conducting the experiment, ensuring its reproducibility in scientific events, and educational laboratories. The scripts developed for the virtual reconstruction have been implemented on \lstinlineMatlab{Matlab} and Jupiter Notebook with documentation available at \url{https://csc1.gitlab.io/archeolab/}. Finally, a training activity have been proposed, which can be adapted to the audience and the goals of the instructor.

\appendix

\section{The mathematical model of X-ray tomography}\label{ref: mathematic}

In this section, the mathematical model of X-ray tomography is described. This model is based on the computerized transverse axial (CTA) scanners, where three-dimensional objects are modeled as a collection of two-dimensional reconstructed slices. As text references for undergraduate students, we suggest \emph{The Mathematics of Medical Imaging} of Timothy Feeman \cite{MR3410524} and \emph{The Radon Transform and Medical Imaging} of Peter Kuchment \cite{MR3137435}. For advanced undergraduate and graduate students, the classical textbook \emph{The Mathematics of Computerized Tomography} of Frank Natterer \cite{MR1847845} is recommended.

The plane $\R^{2}$ is considered with points denoted by the coordinates $(x,y)$. The object is described by a bounded, connected subset of the plane, denoted by $\Omega$. The X-ray attenuation coefficient is modeled by a function $f:\R^{2}\rightarrow \R$, $(x,y)\mapsto f(x,y)$, which is compactly supported in $\Omega$. In this setting, it is assumed that the geometry of the experiment is the parallel beam geometry (i.e., each source has a corresponding detector that measures the attenuation of the emitted beam, and all the beam lines are parallel). In modern application, the fan beam geometry is more frequently employed than the parallel beam geometry. In this case, a source emits several X-ray beams in different directions contained in a plane, and an array of detectors registers their variations in intensity.

The intensity of the X-ray beam, denoted by $I$, is defined as the quantity of radiation per unit area, the latter measured on a plane perpendicular to the X-ray beam. The source point is denoted by $P_0=(x_0,y_0)$ and the detector point by $P_1=(x_1,y_1)$. Let $L$ be the line through $P_0$ and $P_1$ with parametrization $\gamma$ given by
\[
\gamma(t)=(x_0,y_0)+t(\cos\theta,\sin\theta),
\]
where $\theta\in [0,2\pi)$ and $t\in \R$. Here, $S^{1}$ denotes the unit sphere in $\R^{2}$ and is intended as the set of all unitary directions. Conventionally, the center of rotation is the origin $(0,0)$, and sources and detectors rotate around it.

\begin{figure}[h]
	\centering
	\includegraphics[width=.3\linewidth]{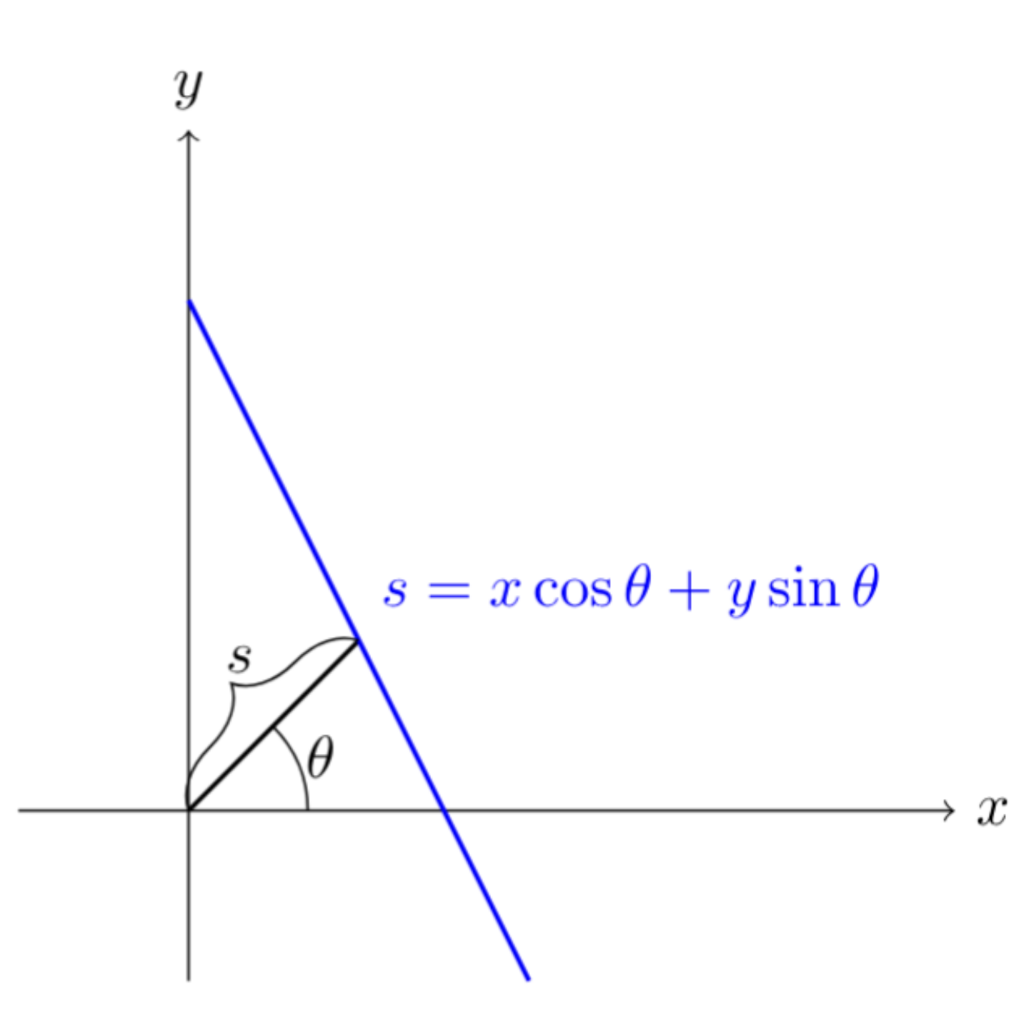}
	\caption{Normal parametrization of lines.}
	\label{fig:normal_eq}
\end{figure}

The physical law that enables to combine the information about the intensity variation of the radiation and the unknown attenuation coefficient $f$ is known as the Beer-Lambert law. This law states that the intensity variation at a point of the object lying on the X-ray beam, denoted by $\gamma(t)$, is directly proportional to the X-ray intensity at a specific point and path length. The standard formula for this law is expressed as follows:
\begin{equation}\label{eqn: beers}
	I(t+\Delta t) - I(t) \simeq - f(t) I(t) \Delta t.
\end{equation}
The function $f(t)=f(\gamma(t))$ represents the attenuation coefficient at position $\gamma(t)$. If the limit of the parameter increment $\Delta t$ goes to zero in equation \eqref{eqn: beers}, under the assumption that the initial intensity, denoted by $I_0$, at the source position $P_0$ for $t=t_0$ is known, the following initial value problem is obtained:
\[
\begin{cases}
	\frac{dI}{dt}(t) = -f(t) I(t) &t\in \R,\\
	I(t_0) = I_0.
\end{cases}
\]
The solution $I$ to this problem is given by the exponential function
\begin{equation}\label{eqn: intensity}
	I = I_0 \exp\Big(-\int_L f d\call\Big),
\end{equation}
where $d\call$ is the arc length element and $L$ is the X-ray beam. Equation \eqref{eqn: intensity} is consistent with the physical experiment, since the intensity $I(t)$ at a position $P_1$ for $t=t_1$ is always smaller or equal to the initial intensity $I_0$. Applying the logarithm on \eqref{eqn: intensity}, and denoting with $I_1=I(t_1)$ the recorded intensity at position $P_1$, the X-ray projection formula is obtained:
\begin{equation}\label{eqn: projection}
	p(t_1) = -\log\bigg(\frac{I_1}{I_0} \bigg) = \int_L f(x,y) d\call. 
\end{equation}
In the plane the projection is formally known as the \textit{Radon Transform}. It is defined as the line integral of the attenuation coefficient $f$ along an X-ray beam $L$:
\begin{equation}
	\RR f(L) = \int_{L} f(x,y) d\call.
\end{equation}
To represent lines, the so-called \textit{normal parametrization} is introduced, defined as follows: for $t\in \R$,
\begin{align*}
	\gamma(t) = (s\cos\theta-t\sin\theta,s\sin\theta + t\cos\theta)
\end{align*}
(see Figure \ref{fig:normal_eq}). Here, the variable $s$ is called affine parameter, denoting the oriented distance of the line $L$ from the origin, while the variable $\theta$ is the counter-clockwise rotation of the line $L$ with respect to the $x$-axis. A useful expression that connects the variables $x, y, s, \theta$ is the following formula, called the normal equation:
\begin{equation}\label{eqn: normal}
	s = x\cos\theta + y\sin \theta.
\end{equation}

Under the normal parametrization, the Radon transform in the plane of a line $L$ characterized by the parameters $(s,\theta)$ corresponds to the projection $p(s,\theta)$. It has the formula
\begin{align*}
	p(s,\theta) &= \int_{-\infty}^{+\infty} f(\gamma(t)) |\gamma'(t)| dt\\ 
	&= \int_{-\infty}^{+\infty} f(s\cos\theta-t\sin\theta,s\sin\theta + t\cos\theta) dt,
\end{align*}
and
\begin{equation}
	\RR f(s,\theta) = p(s,\theta).
\end{equation}
In dimension higher than two, the Radon transform integrates along hyperplanes and not lines (see \cite{MR1847845} for more about the X-ray transform). A pivotal quantity associated with the Radon transform is the \textit{sinogram}, representing the projection data at each angle, denoted by the variable $\theta$, and distance from the origin, denoted by the variable $s$. The name "sinogram" derives from the observation that a fixed point of the object traces out a sinusoidal curve in the sinogram.

The objective is to recover the function $f$ from the knowledge of the projections $p(s,\theta)$, subject to the variation of the parameters $(s,\theta)$. The back projection method can be regarded as a first naive reconstruction technique. This method involves back-projecting each projection onto the plane to recover the value of the function $f$ at a specific point $(x,y)$. This operation gives rise to the adjoint Radon transform, also known as the \textit{back-projection operator}:
\begin{equation}\label{eqn: backoperator}
	\mathcal{B}[\RR f(s,\theta)](x,y) = \int_0^{2\pi} \RR f(s,\theta) d\theta,
\end{equation}
that can be equivalently express in terms of the projection:
\begin{equation}
		\mathcal{B}[p(s,\theta)](x,y) =\int_0^{2\pi} \int_{-\infty}^{+\infty} f(s\cos\theta-t\sin\theta,s\sin\theta + t\cos\theta) dt d\theta.
\end{equation}
Here, the definition from reference \cite[Equ. (2.5)]{MR2207138} is used. The inner integral determines the projection $p(s,\theta)$, which corresponds to a unique point $(s,\theta)$ in the sinogram. The back-projection operation involves the summation of ray-projections over all ray-paths through a given point $(x,y)$. The Back-Projection technique was originally introduced by Oldendorf \cite{Old61} in the context of X-ray tomography. However, the reconstruction of the function $f$ is not optimal due to the fact that the summation of projections is applied not only to points of high density, but also to other points of the object. This results in the presence of star artifacts in the reconstruction and a blurred background. Furthermore, the Back-Projection method smoothens the edges. Examples can be found in \cite{MR3410524}.

The Back-Projection formula can be improved by integrating a filtering operation prior to back-projected. This approach, called \textit{Filtered Back-Projection}, involves the application of a filter to the projections, enabling the enhancement or reduction of high frequencies in the Fourier domain, which are known to be associated with edge detection. 

Let us begin with the definition of the two-dimensional Fourier transform of the function $f(x,y)$ given by
\begin{equation}\label{eqn: inverse-fourier}
	\mathcal{F}f(k_x,k_y) = \frac{1}{2\pi} \int_{\R^{2}} f(x,y)e^{-i(k_x x+k_y y)} dx dy.
\end{equation}

The reconstruction formula presented here is based on the application of the Fourier Slice Theorem, which states that the one-dimensional Fourier transform of the Radon transform, the projection, with respect to the affine parameter $s$ is equal to the two-dimensional Fourier transform of the function $f$ \cite{MR2207138}:
\begin{equation}\label{eqn: slicethem}
	\mathcal{F}f(\sigma \theta) = \frac{1}{\sqrt{2\pi}} \mathcal{F}_s \RR f(\sigma,\theta),
\end{equation}
where
\begin{equation}
	\mathcal{F}_s \RR f(\sigma,\theta) = \frac{1}{\sqrt{2\pi}} \int_{\R} e^{-is\sigma} \RR f(s,\theta) ds.
\end{equation}
To derive the the Filtered Back-Projection formula, commence with representing the attenuation coefficient $f$ in terms of its 2-D Fourier transform:
\begin{equation}\label{eqn: fourierintegral}
		f(x,y) = \frac{1}{2\pi}\int_{\R^{2}} \mathcal{F}f(k_x,k_y)\cdot e^{ i(k_x x+k_y y)} dk_x dk_y.
\end{equation}
In this representation, the function $f$ is expanded as a superposition of sinusoidal waves, encoded in the complex exponential. The parameters $k_x$ and $k_y$ represent the Fourier parameters along the directions $x$ and $y$, respectively. 
By switching to polar coordinates in equation \eqref{eqn: fourierintegral}, thanks to the symmetry of the Radon transform, one can derive
\begin{equation}\label{eqn: intermediate2}
		f(x,y) = \frac{1}{2(2\pi)} \int_0^{2\pi} \int_{\R} \mathcal{F}f(\sigma\theta)\cdot e^{i \sigma(x\cos\theta+y\sin\theta)}|\sigma| d\sigma d\theta.
\end{equation}
Since the value of $k$ ranges from $-\infty$ to $+\infty$, the inner integral has the form of a one-dimensional Fourier transform. Hence, by \eqref{eqn: slicethem} and \eqref{eqn: normal} 
\begin{equation}\label{eqn: intermediate3}
		f(x,y) = \frac{1}{4\pi} \int_0^{2\pi} \frac{1}{\sqrt{2\pi}}\int_{\R} \mathcal{F}_s \RR f(\sigma,\theta)  \cdot e^{i \sigma s}|\sigma| d\sigma d\theta.
\end{equation}
At this point, the inner integral is the inverse Fourier transform $\mathcal{F}^{-1}_{\sigma}$ of the function $\mathcal{F}_s \RR f(\sigma,\theta)|\sigma|$ with respect to the parameter $\sigma$, while the outer integral is the back-projector operator \eqref{eqn: backoperator}. Therefore, a formula for the Filtered Back-Projection is derived:

\begin{equation}
	f(x,y) = \frac{1}{4\pi} \mathcal{B}(\mathcal{F}^{-1}_{\sigma} (|\sigma|\mathcal{F}_s \RR f(\sigma,\theta))).
\end{equation}
The function $|\sigma|$ is referred to as a ramp filter. In practice, the acquired data is subject to noise, which typically exhibits a high-frequency component in the Fourier domain. Consequently, it is common to apply a low-pass filter to the ramp filter, a function that attenuates the high-frequency contributions. 

In the experiment, the Ram-Lak filter was used. This filter was first introduced by G. N. Ramachandran and A. V. Lakshminarayanan in 1971 \cite{MR287750}. This type of filter is applied to the ramp filter in the Fourier domain yielding the function $|\sigma|$ if $\sigma\leq 1$, and $0$ otherwise (see also \cite{MR3410524}).

\section*{Acknowledgements}

This research was funded in whole or in part by the Austrian Science Fund (FWF) SFB 10.55776/F68 ``Tomography Across the Scales'', project F6801. For open access purposes, the authors have applied a CC BY public copyright license to any author-accepted manuscript version arising from this submission. The financial support by the Austrian Federal Ministry for Digital and Economic Affairs, the National Foundation for Research, Technology and Development and the Christian Doppler Research Association is gratefully acknowledged.

\section*{References}
\renewcommand{\i}{\ii}
\printbibliography[heading=none]

\end{document}